\documentclass[11pt,a4paper]{article}

\usepackage{amssymb}

\def\smallcirc{{\raise 0.5pt \hbox{$\scriptstyle\circ$}}}
\def\smallover#1/#2{\hbox{$\textstyle{#1\over#2}$}} %

\let\ssection=\section
\renewcommand{\section}{\setcounter{equation}{0}\ssection}

\title{Supersymmetric Chaplygin gas}
\bigskip
\author{Mokhtar ~Hassa\"{\i}ne\\
Universit\'e de
Tours\\
Laboratoire de Math\'ematiques et de Physique Th\'eorique,\\
Parc
de Grandmont, 37200 Tours,
France}
\begin{document}
\maketitle
\begin{abstract}
Using a Kaluza-Klein framework, we consider a relativistic fluid whose projection yields the supersymmetric non-relativistic Chaplygin gas introduced by Bergner-Jackiw-Polychronakos and by Hoppe. The conserved (super)charges of the Chaplygin gas are obtained as the projection of those arising in the extended model.
\end{abstract}
\section{Introduction}
Recently, the strange properties of an isentropic fluid (called Chaplygin gas) arising from membrane theory have attracted much attention \cite{JP}-\cite{HH}. This gas is described by the following equations,
\begin{eqnarray}
\begin{array}{l}
\partial_tR+\vec{\nabla}\cdot(R\,\vec{v})=0\\
[0.5mm]\\
\partial_t\vec{v}+(\vec{v}\cdot\vec{\nabla})\,\vec{v}=-\displaystyle{\frac{1}{R}\vec{\nabla}p}
\end{array}
\label{1}
\end{eqnarray}
where $R$ is the density of the fluid, $\vec{v}=\vec{\nabla}\Theta$ is the irrotational velocity and $p$ represents the pressure. This fluid is characterised by a negative pressure $-2\lambda/R$ ($\lambda$ is a positif constant). The supersymmetric generalization of the planar Chaplygin gas arising from a supermembrane theory in $(3+1)$-dimensions was proposed in \cite{JP} and \cite{HOP}. Jackiw and Polychronakos \cite{JP}, in particular introduced anticommuting Grassmann variables in order to get a non-irrotational velocity (the vorticity is indeed generated by the fermion fields).
Following the same construction, Bergner and Jackiw \cite{BJ} have
obtained an integrable supersymmetric fluid model in
$(1+1)$-dimensions with large symmetries, see Section $3$. Similar
remarks were obtained by Bazeia in the purely bosonic model
\cite{BB}. Let us mention that an another interesting gas model,
namely the polytropic gas, has been studied by Das and Popowicz
\cite{DAS}. They analyse the differences and the similarities between
the supersymmetric polytropic gas and the supersymmetric Chaplygin gas. 

In this letter, we propose an original method to construct the conserved (super)charges of the planar supersymmetric and non-relativistic Chaplygin gas \cite{JP}. In this letter, we consider a relativistic perfect fluid on an extended space whose projection gives the planar model. In order to include supersymmetry, we generalize the usual approach to relativistic perfect fluids \cite{LL}. Indeed, our relativistic fluid is governed by two continuity equations (instead of one, as usual) and by the conservation law of the energy momentum tensor. The second continuity equation is a fermionic equation which is absent when the theory is purely bosonic. We prove a correspondance between this relativistic fluid and the supersymmetric Chaplygin model and we use this correspondance to construct the conserved (super)charges. In the last section, we show similar results in the case of the lineal model.
\section{The planar Chaplygin gas}
The equations of motion of the supersymmetric planar Chaplygin gas are
\begin{eqnarray}
\begin{array}{l}
\partial_t R+\vec{\nabla}\cdot(R\vec{v})=0,\\
[0.5mm]\\
\displaystyle{\partial_t\psi+\vec{v}\cdot\vec{\nabla}\psi=\frac{\sqrt{2\lambda}}{R}\,\vec{\alpha}\cdot\vec{\nabla}\psi},\\
[0.5mm]\\
\displaystyle{\partial_t\vec{v}+(\vec{v}\cdot\vec{\nabla})\vec{v}=
-\frac{1}{R}\vec{\nabla}\left(\frac{-2\lambda}{R}\right)+\frac{\sqrt{2\lambda}}{R}\,\vec{\nabla}\psi\,\vec{\alpha}\cdot\vec{\nabla}\psi}.
\end{array}
\label{2}
\end{eqnarray}
The velocity is defined as 
\begin{eqnarray}
\vec{v}=\vec{\nabla}\Theta-\frac{1}{2}\psi\,\vec{\nabla}\psi,
\label{clebsch}
\end{eqnarray}
which provides a Clebsch formula \cite{DESER} for $\vec{v}$. The Grassmann variables $\psi$ are Majorana spinors and the matrices $\alpha^i$ ($i=1, 2$) are given in terms of the matrices Pauli by $\alpha^1=\sigma^1$ and $\alpha^2=\sigma^3$. The system (\ref{2}) can be derived from the following Lagrangian
\begin{eqnarray}
L=-R\left(\partial_t\Theta-\frac{1}{2}\psi\partial_t\psi\right)-\frac{R}{2}\left(\vec{\nabla}\Theta-\frac{1}{2}\psi\vec{\nabla}\psi\right)^2-\frac{\lambda}{R}-\frac{\sqrt{2\lambda}}{2}\psi\,\vec{\alpha}\cdot\vec{\nabla}\psi
\label{3}
\end{eqnarray}
In \cite{JP}, Jackiw and Polychronakos showed that this model admits $4$ fermionic supercharges given by
\begin{eqnarray}
Q_j=\int d^2x\left(R\,\vec{v}\cdot(\vec{\alpha}_{jk}\,\psi_k)+\sqrt{2\lambda}\,\psi_j\right)\qquad\quad j=1, 2,
\label{4}
\end{eqnarray}
\begin{eqnarray}
\begin{array}{lcccccccccccccc}
\tilde{Q}_j=\int d^2x\,R\,\psi_j\qquad & & & & & & & & & & & & j=1, 2,
\label{5}
\end{array}
\end{eqnarray}
where $\vec{\alpha}_{jk}$ are the components of the $\vec{\alpha}$ matrices.

We propose now a relativistic fluid model whose projection leads to the previous one. Following \cite{DBKP}, let us consider a Lorentz $4$-manifold $(M,g)$ which is endowed with a covariantly constant and lightlike vector field, $\xi$. The coordinates on $M$ are $(\vec{x},t,s)$ where $s$ is an additional ``vertical'' coordinate. The quotient of $M$ by the integral curves of $\xi$ is a $(2+1)$ non-relativistic space-time \cite{DBKP}. In our case, we choose the flat Minkowski metric written in light-cone coordinates as $d\vec{x}^2+2dtds$, and $\xi^\mu\partial_\mu=\partial_s$. In $M$, we consider the bosonic fields $\rho$ and $\theta$ and the fermionic Majorana spinors $\tilde{\psi}$. In what follows, the Greek letters $\mu, \nu\cdots=\vec{x}, t, s$ are indices on the extended space and the indices $j$ and $k$ run over the spatial components. 

On the extended space $M$, we consider a relativistic Lagrangian which describes the motion of a perfect fluid and whose projection (as we shall see below) leads to the non-relativistic one (\ref{3}). This relativistic Lagrangian is the following,
\begin{eqnarray}
{\cal L}=-\frac{\rho}{2}\,v_\mu\,v^\mu-\frac{\lambda}{\rho}-\frac{\sqrt{2\lambda}}{2}\,\Psi^\dagger\left[A\,\gamma^j\partial_j\Psi\right],
\label{3'}
\end{eqnarray}
where $\Psi$ is a $4$-spinor with components $(0,0,\psi_1,\psi_2)$ and where $\psi_k$ are Majorana spinors which do not depend on the variable $s$. The $4$-velocity $v^\mu$ is defined as 
$$
v^\mu=\partial^\mu\theta-\frac{1}{2}\psi\partial^\mu\psi.
$$
The Lagrangian (\ref{3'}) involves also the $\gamma$ matrices whose representation in the light-cone coordinates is given by
$$
\begin{array}{cc}
\gamma^t=\left(
\begin{array}{cc}
0 & \begin{array}{cc}
    0 & 0 \\
    0 & -\sqrt{2}
    \end{array} 
\\
 \begin{array}{cc}
    -\sqrt{2} & 0 \\
    0 & 0
    \end{array} & 0
\end{array}
\right)
& 
\gamma^s=\left(
\begin{array}{cc}
0 & \begin{array}{cc}
    \sqrt{2} &  0\\
    0 &  0
    \end{array}
\\
 \begin{array}{cc}
    0 & 0 \\
    0 & \sqrt{2}
    \end{array} & 0
\end{array}
\right)
\end{array}
$$
and 
$$
\gamma^k=\left(
\begin{array}{cc}
0 & \alpha_k\\
-\alpha_k & 0
\end{array}
\right)\qquad\quad\mbox{for}\,k=1, 2
$$
The $\gamma$ matrices satisfy $\{ \gamma^\mu,\gamma^\nu\}= -2\,g^{\mu\nu}$ where $g_{\mu\nu}$ is the flat Minkowski metric $d\vec{x}^2+2dtds$. The $4\times 4$ matrix is
$$
A=\left(
\begin{array}{cc}
0 & -I_2\\
-I_2 & 0
\end{array}
\right)=A^{-1}.
$$
The variation of (\ref{3'}) with respect to $\theta$ leads to the equation
\begin{eqnarray}
\partial_\mu(\rho\,v^\mu)=0,
\label{continuity}
\end{eqnarray}
called bosonic continuity equation (this equation is indeed present even if the fermionic field is absent). The equations associated to $\psi$ read
\begin{eqnarray}
v^\mu\partial_\mu\Psi=\frac{\sqrt{2\lambda}}{\rho}\, A\,\gamma^k\partial_k\Psi.
\label{gammaequation}
\end{eqnarray}
Finally, the variation of (\ref{3'}) with respect to $\rho$ yields to 
\begin{eqnarray}
2\lambda=\rho^2\,v^\mu\,v_\mu. 
\label{relax}
\end{eqnarray}
This last equation with (\ref{gammaequation}) lead to an Euler-type equation,
\begin{eqnarray}
\displaystyle{v^\mu\partial_\mu\,v_k=-\frac{1}{\rho}\partial_k\left(\frac{-2\lambda}{\rho}\right)+\frac{\sqrt{2\lambda}}{\rho}(\partial_k\Psi)^\dagger \left[A\,\gamma^j\partial_j\Psi\right]}.
\label{eulerequation}
\end{eqnarray}
We argue that our system (\ref{continuity}-\ref{eulerequation}) describes a relativistic fluid. Following \cite{LL}, a relativistic fluid can be described by a continuity equation, augmented by a conserved energy momentum tensor. In our case, we generalize this construction by considering two continuity equations (rather than one) to include the supersymmetry. Indeed, using the bosonic continuity equation (\ref{continuity}), the second equation of our relativistic system (\ref{gammaequation}) is in fact a continuity equation,
\begin{eqnarray}
\partial_\mu\left(\rho\,v^\mu\,\Psi\right)=\sqrt{2\lambda}\, A\,\gamma^k\partial_k\Psi.
\label{gammaequation'}
\end{eqnarray}
In fact, the right term is a total spatial divergence and, contrary to (\ref{continuity}) this equation (called fermionic continuity equation) is meaningless when the theory is purely bosonic. Varying now the relativistic Lagrangian (\ref{3'}) with respect to the metric, we obtain a symmetric energy-momentum tensor, namely
\begin{eqnarray}
\vartheta_{\mu\nu}=-g_{\mu\nu}\,{\cal L}-\rho\,v_\mu\,v_\nu+\frac{\sqrt{2\lambda}}{4}\,\delta_\mu^j\,\delta_\nu^k\left(\Psi^\dagger A\left[\gamma_j\partial_k\Psi+\gamma_k\partial_j\Psi\right]\right).
\label{tensor}
\end{eqnarray}
This tensor is conserved, $\partial^\mu \vartheta_{\mu\nu}=0$, and the spatial component of this relation gives the relativistic Euler equation (\ref{eulerequation}). If we introduce the pressure $p=-2\lambda/\rho$ and the unitary velocity $u^\mu$ defined by $n\,u^\mu=\rho\,v^\mu$ (so that $n$ is proportional to the proper particle density ), then the expression (\ref{tensor}) becomes
\begin{eqnarray}
\begin{array}{lll}
\vartheta_{\mu\nu}=& - & \displaystyle{g_{\mu\nu}\,p-\frac{n^2}{\rho}\,u_\mu\,u_\nu+\frac{\sqrt{2\lambda}}{2}\,g_{\mu\nu}\left(\Psi^\dagger A\,\gamma^j\partial_j\Psi\right)}\\
[4mm]
                   & + &\displaystyle{\frac{\sqrt{2\lambda}}{4}\,\delta_\mu^j\,\delta_\nu^k\left(\Psi^\dagger A\left[\gamma_j\partial_k\Psi+\gamma_k\partial_j\Psi\right]\right)}.
\label{tensor'}
\end{array}
\end{eqnarray}
Curiously, this tensor is the sum of two separed conserved tensors,
\begin{eqnarray}
\vartheta_{\mu\nu}=T_{\mu\nu}+\Omega_{\mu\nu}.
\label{ro+}
\end{eqnarray}
The first conserved tensor given by
\begin{eqnarray}
T_{\mu\nu}=-g_{\mu\nu}\,p-\frac{n^2}{\rho}u_\mu\,u_\nu,\qquad \partial_\mu T^{\mu\nu}=0.
\label{man}
\end{eqnarray}
describes a perfect fluid \cite{LL} (this tensor is present even in the purely bosonic model). The second tensor $\Omega_{\mu\nu}$ interpreted as an improvement term \cite{JP} and \cite{VH} is zero when the fermionic field is absent.

Now, if the relativistic fields $\rho$, $\theta$ and $\tilde{\psi}$ are related to the non-relativistic ones, $R$, $\Theta$ and $\psi$ by the equivariance condition\cite{HH} and \cite{DBKP}, namely by
\begin{eqnarray}
\rho=R(t, \vec{x})\quad\mbox{and}\quad \theta=\Theta(t, \vec{x})+s,
\label{equivariance}
\end{eqnarray}
then the equations (\ref{continuity}-\ref{eulerequation}) project to the non-relativistic ones (\ref{2}). Similarly, the Lagrangian (\ref{3'}) becomes exactly (\ref{3}). This correspondance is very useful to obtain the conserved quantities of the non-relativistic model. Indeed, the conserved (super)current will be obtained as the projection of these conservation laws. We distinguish three types of charges :\\
\\
$\bullet$ \underline{The extended Galileo charges:} Our relativistic system is described by the Lagrangian (\ref{3'}) augmented by the equivariance condition (\ref{equivariance}). The isometries of the metric $g$ are symmetries of the Lagrangian (\ref{3'}). In fact, under an isometry transformation the velocity $v^\mu$ and the fermionic field $\psi$ are invariant. However, because of the equivariance condition (\ref{equivariance}), we restrict ourselves to isometries which preserve the vertical vector. These $\xi$-preserving isometries are given by
\begin{eqnarray}
X^\mu\partial_\mu=\epsilon\partial_t+(\vec{\gamma}+t\vec{\beta})\cdot\vec{\nabla}+(\eta-\beta 
x)\partial_s
\label{isometries'}
\end{eqnarray} 
where $\epsilon$, $\vec{\gamma}$, $\vec{\beta}$ and $\eta$ are the infinitesimal parameters associated (respectively) to the time translation, space translation, boost and vertical translation. For any isometry which satisfies (\ref{isometries'}), the $4$-vector field $k^\mu=\vartheta^{\mu}_{\ \nu}\,X^\nu$ is conserved and $s$-independent. Consequently, this conserved $4$-vector projects under equivariance to a conserved $3$-current whose temporal component,
\begin{eqnarray}
\int d\vec{x}\, k^t=\int d\vec{x}\,g^t_{\ \mu}\,\left(\sqrt{2\lambda}{\rho}+\frac{\sqrt{2\lambda}}{2}\Psi^\dagger\,A\,\gamma^j\partial_j\Psi\right)X^\mu-\rho\,v_\mu\,X^\mu
\label{galileo}
\end{eqnarray} 
does not depend on time. The quantities obtained with each isometries (\ref{isometries'}) form the extended Galileo group.\\
\\
$\bullet$ \underline{The Poincar\'e charges:} Previously, we have seen that because of the equivariance condition (\ref{equivariance}), we consider only $\xi-$preserving isometries. Motived by our previouys paper \cite{HH}, we can relax this too restrictive condition. Indeed, let us suppose that the fields $\rho$ and $v^\mu$ depend explicitly on the variable $s$ and, they are related to the non-relativistic field $R$ and $\Theta$ by the new condition,
\begin{eqnarray}
\left\lbrace
\begin{array}{l}
v^\mu\left(t,\vec{x},-\Theta(t,\vec{x})\right)=\left[\partial^\mu\Theta(t,\vec{x})-\frac{1}{2}\psi\partial^\mu\psi\right]\partial_s\theta\vert_{\left(t,\vec{x},-\Theta(t,\vec{x})\right)}\\
\\
R(t,\vec{x})=\rho\left(t,\vec{x},-\Theta(t,\vec{x})\right)
\partial_s\theta\vert_{\left(t,\vec{x},-\Theta(t,\vec{x})\right)}.
\label{fieldtransformhh}
\end{array}
\right.
\end{eqnarray}
Given any solutions of equations (\ref{continuity}-\ref{relax}), the first relation provides a way to construcut the field $\Theta$ solution of the non-relativistic equations (\ref{2}). The second relation gives the correspondance between the fields $\rho$ and $R$. Let us point out that in the case of the classical equivariance (i.e. $\partial_s\theta=1$), these relations lead to the classical ones (\ref{equivariance}). Using this new equivariance (\ref{fieldtransformhh}), we prove easily that equations (\ref{continuity}-\ref{relax}) project to the non-relativistic equations (\ref{2}). Consequently, the isometries (which not preserve necessarily the vector field $\xi$) are symmetries of the system defined by equations (\ref{continuity}-\ref{relax}) augmented by the condition (\ref{fieldtransformhh}). The $\xi$ non-preserving isometries are
\begin{eqnarray}
X^\mu\partial_\mu=(d\,t+\vec{\omega}\cdot\vec{x})\partial_t-s\,\vec{\omega}\cdot\vec{\nabla}-d\,s\,\partial_s,
\label{isometries''}
\end{eqnarray} 
where $d$ and $\vec{\omega}$ are respectively the paramaters associated to the time dilatation and the anti-boosts \cite{HH}. We can not project, as previously, the $4$ conserved current $k^\mu$ because of its explicit dependence on the additional variable $s$. However, using the Euler equation (\ref{eulerequation}), we show that the vector $k^\mu$ taken in the particular point $(t,\vec{x},-\Theta(t,\vec{x}))$ is always conserved \cite{HH}. A similar calculation (\ref{galileo}) leads to the following conserved quantities ,
\begin{eqnarray}
\begin{array}{ccc}
D=\displaystyle{\int \left(t\,{\cal H}-R\,\Theta\right)\,d\vec{x}} & &\mbox{time dilatation},\\
\\
\vec{G}=\displaystyle{\int \left(\vec{x}\,{\cal H}-\Theta\vec{{\cal P}}\right)\,d\vec{x}} & &\mbox{anti-boosts},\\
\end{array}
\label{poincare}
\end{eqnarray}
where ${\cal H}$ and $\vec{{\cal P}}$ are respectively the energy density and the momentum density. These $3$ charges with the extended Galileo ones form the Poincar\'e group in $(3+1)$ dimensions.\\
\\
$\bullet$ \underline{The supercharges:} First, the fermionic continuity equation (\ref{gammaequation'}) gives a conserved supercurrent on $M$ wich does not depend on the additional variable $s$. Its temporal component yields the conserved quantity (\ref{4}). Moreover, combining the equations (\ref{continuity}) and (\ref{eulerequation}), we have
$$
\partial_\mu\left(\rho\,v^\mu\,v_k\right)-\partial_k(\frac{2\lambda}{\rho})=\sqrt{2\lambda}\,(\partial_k\Psi)^\dagger \left(A\gamma^j\partial_j\Psi\right).
$$
Contracting this expression with $\gamma^k\Psi$, we obtain (after integration) 
\begin{eqnarray}
\begin{array}{lll}
\partial_\mu\left(\rho\,v^\mu\,v_k\,\gamma^k\Psi\right)-\displaystyle{\partial_k(\frac{2\lambda}{\rho}\,\gamma^k\Psi)}& = &\sqrt{2\lambda}\,A\left(v^\mu\partial_\mu\Psi -v_k\gamma^k\gamma^j\partial_j\Psi\right) \\
[4mm]
& & +\sqrt{2\lambda}\,(\partial_k\Psi)^\dagger\left(A\gamma^j\partial_j\Psi\right)\gamma^k\Psi.
\end{array}
\label{ley}
\end{eqnarray}
Let us remark that only the two first components of this expression are non zero. After a tedious calculation, we prove that the first term on the right is equal to
$$
\sqrt{2\lambda}\,\partial_t\psi_k-\epsilon^{km}\left[\epsilon^{ij}\partial_i(v_j\psi_m)\right]+\omega\,\epsilon^{km}\,\psi_m,
$$
where $\omega$ is the vorticity, $\omega=\epsilon^{ij}\partial_i\,v_j=-1/2\,\epsilon^{ij}\partial_i\psi\partial_j\psi$. Finally, using this relation, the term in right in (\ref{ley}) becomes
$$
\sqrt{2\lambda}\,\partial_t\psi_k-\epsilon^{km}\left[\epsilon^{ij}\partial_i(v_j\psi_m)\right]+\epsilon^{ij}\partial_i\left(\psi_2\,(\partial_j \psi_k)\,\psi_1\right).
$$
Consequently, we have proved that the relation (\ref{ley}) is a conservation law whose temporal component projects to the density of the supercharge (\ref{5}).
\section{The lineal Chaplygin gas}
The lineal version of the Chaplygin gas is governed by the following equations
\begin{eqnarray}
\left\{
\begin{array}{c}
\partial_t R+\partial_x(R v)=0\\
[3mm]
\displaystyle{\partial_t\psi+\left(v+\frac{\sqrt{2\lambda}}{R}\right)\partial_x\psi=0}\\
\displaystyle{\partial_t v+v\partial_x v=-\frac{1}{R}\partial_x\left(\frac{-2\lambda}{R}\right)}.
\end{array}
\right.
\label{eulernonre}
\end{eqnarray}
This model is more particular than the planar one. Indeed, in \cite{BJ}, it was shown that this model is completely integrable, and admits an infinite ladder of ``bosonic''conserved charges, namely
\begin{eqnarray}
\displaystyle{Q_n^{\pm}=\frac{1}{n!}\int R\left(v\pm\frac{\sqrt{2\lambda}}{R}\right)^n dx}\qquad\mbox{for}\,\,n=0, 1\cdots
\label{infinitebosonic}
\end{eqnarray}
The first three terms of the sequence $Q_n^+$ give some of the generators of the extended Galileo group (namely the particle number $N$, the momentum $P$ and the energy $H$). Curiously, the boosts, $B=t P-\int x\,R$ as well as the Poincar\'e quantitites (\ref{poincare}) which are also conserved do not appear in the sequence $Q_n^+$. The supersymmetric model also admits an infinite ladder of fermionic supercharges \cite{BJ} given by
\begin{eqnarray}
\displaystyle{\tilde{Q}_n=\int R\left(v-\frac{\sqrt{2\lambda}}{R}\right)^n\psi\, dx}\qquad \mbox{for}\,\,n=0, 1\cdots
\label{infinitefermionic}
\end{eqnarray}

Similarly, we consider a relativistic fluid in $(2+1)$ dimensions  whose motion is governed by :\\
$\bullet$ two continuity equations (one bosonic and the other fermionic)
\begin{eqnarray}
\left\lbrace\begin{array}{l}
\partial_\mu(\rho\,v^\mu)=0\\
\\
\partial_\mu\left[\rho\,(v^\mu-V^\mu)\psi\right]=0.
\end{array}
\right.
\label{bosonicequa}
\end{eqnarray}
Here $V^\mu$ is a spacelike vector
\begin{eqnarray}
V^\mu\partial_{\mu}=-\frac{\sqrt{2\lambda}}{\rho}\partial_x.
\label{Vmu}
\end{eqnarray}
$\bullet$ a conserved energy momentum tensor,
\begin{eqnarray}
\vartheta_{\mu\nu}=\underbrace{-g_{\mu\nu}\,p-\frac{n^2}{\rho}u_\mu\,u_\nu}_{T_{\mu\nu}}+\underbrace{\displaystyle{\sqrt{2\lambda}\,
g_{\mu\nu}\, v+n\,u_\mu\,V_\nu+n\,u_\nu\,V_\mu-\rho\,V_\mu\,V_\nu}}_{\Omega_{\mu\nu}}.
\label{tensorr}
\end{eqnarray}
This tensor is also the sum of two separated conserved tensors, $T_{\mu\nu}$ and $\Omega_{\mu\nu}$. If we suppose now that the fluid is isentropic with a negative pressure, $p=-2\lambda/\rho$, then the conservation of $T^{\mu\nu}$ leads to
\begin{eqnarray}
v^\mu\partial_\mu\,v_\sigma=-\frac{1}{\rho}\partial_\sigma\left(\frac{-2\lambda}{\rho}\right)\qquad\mbox{for}\quad \sigma=x, t, s.
\label{manu'''}
\end{eqnarray}
As previously, under equivariance condition (\ref{equivariance}), this relativistic fluid (\ref{bosonicequa}-\ref{manu'''}) projects to the lineal non-relativistic model.

Let us now derive the conservation laws for the lineal model. The isometric charges (Galileo and Poincar\'e charges) are obtained as in the planar model. The $3$-vector field $k^\mu=\vartheta^\mu_{\ \nu} X^\nu$ is conserved for all isometries,
\begin{eqnarray}
\partial_\mu k^\mu=\left[g^\mu_{\ \nu}\left(\frac{2\lambda}{\rho}+\sqrt{2\lambda}v\right)-\rho\,(v^\mu-V^\mu)\,(v_\nu-V_\nu)\right]\partial_\mu X^\nu.
\label{lynx}
\end{eqnarray} 
This expression vanishes for all Killing vectors because $\partial_\mu X^\nu$ is proportional to the Lie derivative of $X$.

Moreover, the lineal system admits further conserved quantities which are not associated with Killing vector. Starting from the observation that $\rho V^\mu=-\sqrt{2\lambda}\,\delta^\mu_x$, the first equation in (\ref{bosonicequa}) implies that 
\begin{eqnarray}
\partial_\mu\left(\rho\,(v^\mu\mp V^\mu)\right)=0,
\label{bosonicequa'}
\end{eqnarray}
Using this relation, the second equation in (\ref{bosonicequa}) yields to $(v^\mu-V^\mu)\,\partial_\mu\psi=0$. Consequently, for each $p$ integer we have
\begin{eqnarray}
(v^\mu-V^\mu)\,\partial_\mu\,\psi^p=0.
\label{fermionicequa'}
\end{eqnarray}
Using the same equations, we show first that $(v^\mu\mp V^\mu)\,\partial_\mu\,(v\pm V)=0$ and 
\begin{eqnarray}
(v^\mu\mp V^\mu)\,\partial_\mu\,(v\pm V)^{p}=0.
\label{tensor''}
\end{eqnarray}
Combining equations (\ref{fermionicequa'}) and (\ref{tensor''}), we have for the upper sign
\begin{eqnarray}
(v^\mu-V^\mu)\,\partial_\mu\,\left[(v+V)^n\,\psi^p\right]=0,
\label{total}
\end{eqnarray}
where $n$ and $p$ are integers. We use this last equation to construct non-Killing vectors which satisfy $\partial_\mu k^\mu=0$, (\ref{lynx}). In particular, if we suppose that the vector field $X^\mu$ has the particular form, $X^\mu\partial_\mu=X\partial_s$ (where $X$ is a function which does not depend on the $s$ variable), then the divergence of $k^\mu$ (\ref{lynx}) becomes
$$
\partial_\mu k^\mu=-\rho\,(v^\mu-V^\mu)\,\partial_\mu X.
$$
Solutions of $\partial_\mu k^\mu=0$ are given by $X=(v+V)^n\,\psi^p$, (\ref{total}). The projection of $k^t$ gives the conserved quantity,
\begin{eqnarray}
{\cal Q}_{n}^p=\int dx\;R\,(v-\frac{2\lambda}{R})^n\,\psi^p.
\label{leyleprince}
\end{eqnarray}
If we suppose that the following brackets yield \cite{JP}, 
$$
\{ \psi_a;\psi_b\}=-\frac{\delta_{ab}}{R}\,\delta ,\quad \{ v;\psi\}=-\frac{\partial_x\psi}{2R}\,\delta\quad \mbox{and}\quad \{ v;R\}=-\partial_x\delta 
$$
then the quantities ${\cal Q}_{n,p}$ span a closed algebra,
\begin{eqnarray*}
\{B;{\cal Q}_{n}^p\}=n {\cal Q}_{n-1}^p\quad \{D;{\cal Q}_{n}^p\}=(n-\frac{1}{2}) {\cal Q}_{n}^p,\quad\{G;{\cal Q}_{n}^p\}=(\frac{n}{2}-\frac{1}{2}){\cal Q}_{n-1}^p.
\end{eqnarray*}
The lineal system admits other symmetries which are not associated to the vector $k^\mu$. Indeed, combining the equations (\ref{bosonicequa'}) and (\ref{tensor''}) for the lower sign, we obtain another conserved vector, namely
$$
\tilde{k}^\mu=\rho\,(v^\mu+V^\mu)\,(v-V)^n,
$$
whose temporal component yields to the conserved quantity,
\begin{eqnarray}
\tilde{{\cal Q}}_{n}=\int dx\;R\,(v+\frac{2\lambda}{R})^n.
\label{leyleprince'}
\end{eqnarray}
The algebra verified by (\ref{leyleprince'}) is also a closed algebra with relations,
\begin{eqnarray*}
\{B;\tilde{{\cal Q}}_{n}\}=n \tilde{{\cal Q}}_{n-1},\quad \{D;\tilde{{\cal Q}}_{n}\}=(n-1) \tilde{{\cal Q}}_{n},\quad \{G;\tilde{{\cal Q}}_{n}\}=(\frac{n}{2}-1)\tilde{{\cal Q}}_{n-1}.
\end{eqnarray*}
\\
\noindent
{\bf Acknowledgements} : I am indebted to P. Horv\'athy, J. Hoppe and O. Ley for discussions. It is a pleasure to thank C. Duval, P. Forg\'acs and N. Mohammedi for their fruitful remarks.

\end{document}